\documentclass[12pt,english]{article}
\usepackage[T1]{fontenc}
\usepackage[latin9]{inputenc}
\usepackage{listings}
\usepackage{amsmath}
\usepackage{amssymb}

\makeatletter

\providecommand{\tabularnewline}{\\}

\usepackage{amsfonts}
\providecommand{\U}[1]{\protect\rule{.1in}{.1in}}

\makeatother

\usepackage{babel}

\begin{document}

\title{\textbf{Data Sets of Very Large Linear Feasibility Problems Solved
by Projection Methods}}

\author{Wei Chen\\
Department of Radiation Oncology\\
Massachusetts General Hospital and Harvard Medical School, \\
Boston, MA 02114, USA}

\date{March 2, 2011}
\maketitle
\begin{abstract}
We give a link to a page on the Web on which we deposited a set of
eight huge Linear Programming (LP) problems for Intensity-Modulated
Proton Therapy (IMPT) treatment planning. These huge LP problems were
employed in our recent research and we were asked to make them public.
\end{abstract}

\section{Introduction}

In our recent research \cite{cccdh,ccmzkh} we employed \textit{projection
methods }to solve some huge linear feasibility problems that arise
in Intensity-Modulated Proton Therapy (IMPT) treatment planning. These
huge Linear Programming (LP) problems present a challenge to every
LP solver or solution method and we have made the case in our research
for tackling them with some elaborate projection methods.

Since the papers \cite{cccdh,ccmzkh} are available on the Web we
are not going into any further details about the research itself here.
We only wish to mention that, in the language of \cite{cccdh}: {}``The
main advantage of projection methods, which makes them successful
in many real-world applications, is computational. They have the ability
to handle some huge-size problems of dimensions beyond which more
sophisticated methods cease to be efficient or even applicable due
to memory requirements. This is so because the building blocks of
a projection algorithm are the projections onto the given individual
sets, which are assumed to be easy to perform, and the algorithmic
structure is either sequential or simultaneous, or in-between, as
in the block-iterative projection methods or in the more recent string-averaging
projection methods.\textquotedblright{}

The purpose of this Technical Report, which we do not intend to publish
otherwise, is to announce the availability of our data sets for these
problems along with detailed instructions where to get them and how
to read them. We were asked to post these data sets publicly and we
are pleased to do so.

The data sets are accessible from: \\
 http://dig.cs.gc.cuny.edu/\textasciitilde{}wei/web/?page\_id=221. 

Please contact chen.wei@mgh.harvard.edu for any related questions.

\section{A Set of 8 Huge LP Problems for IMPT}

The following is the information for each LP problem associated with
\textit{Task0, Task1, \ldots{}, Task7}. For details about these tasks
consult the papers where the research is reported \cite{cccdh,ccmzkh}.
The data are binary, double precision, in the order of $a,c,d,e,f,A$
as in\begin{equation}
\min\left\langle a,x\right\rangle \text{ subject to }c\leq Ax\leq d\text{ and }e\leq x\leq f.\end{equation}

The components of the vectors $c,d,e,f$ can be infinite numbers (DBL\_MAX
in double precision in C language). The sparse matrix $A$ is in column
ordered sparse matrix form as described in the MOSEK documentation,
which is standard in storage of huge sparse matrix. For example, it
can be used in CPLEX and other LP solvers as well. The four arrays
of the matrix $A$ are in the order of: $ptrb,ptre,asub,aval$. For
a sample C code of how to read the data see the Appendix below.

\begin{tabular}{|c|c|c|c|c|c|}
\hline 
 & \#Rows & \#Columns & \#NonZeros & Objective & Data\tabularnewline
\hline
\hline 
Task0 & 302,491 & 13,734 & 62,256,376 & MIN & \textasciitilde{}750M\tabularnewline
\hline 
Task1 & 314,546 & 13,734 & 74,554,123 & MAX & \textasciitilde{}900M\tabularnewline
\hline 
Task2 & 302,491 & 13,734 & 62,256,376 & MIN & \textasciitilde{}750M\tabularnewline
\hline 
Task3 & 302,491 & 13,734 & 62,256,376 & MIN & \textasciitilde{}750M\tabularnewline
\hline 
Task4 & 302,491 & 13,734 & 62,256,376 & MIN & \textasciitilde{}750M\tabularnewline
\hline 
Task5 & 302,491 & 13,734 & 62,256,376 & MIN & \textasciitilde{}750M\tabularnewline
\hline 
Task6 & 302,491 & 13,734 & 62,256,376 & MIN & \textasciitilde{}750M\tabularnewline
\hline 
Task7 & 604,982 & 13,734 & 124,754,745 & MIN & \textasciitilde{}1.5G\tabularnewline
\hline
\end{tabular}

\section{Appendix}

The sample code to read the data in C language:
\begin{lstlisting}[basicstyle={\footnotesize\ttfamily},breaklines=true,language=C,showstringspaces=false]
char DataFileName[50] = "./0.dat"; 
FILE*  file = fopen(DataFileName, "rb"); 
if(file == NULL){
cout<<endl<<"**** fail to open file "<<DataFileName<<" ****"<<endl; 
return 0; }else{
cout<<endl<<"Reading a, c, d, e, f and ptrb, ptre, asub, aval of A from the binary file "<<DataFileName<<"...";}
fread(a, sizeof(double), numCol, file); 
fread(c, sizeof(double), numRow, file); 
fread(d, sizeof(double), numRow, file); 
fread(e, sizeof(double), numCol, file); 
fread(f, sizeof(double), numCol, file); 
fread(ptrb, sizeof(int), numRow, file); 
fread(ptre, sizeof(int), numRow, file); 
fread(asub, sizeof(int), numNonZeros, file); 
fread(aval, sizeof(double), numNonZeros, file);
fclose(file);
\end{lstlisting}

\end{document}